\documentclass[nofootinbib, twocolumn, floatfix]{revtex4-1}

\usepackage{graphicx}

\begin{document}

\title{Upper limits on the probability of an interstellar civilization arising in the local Solar neighborhood}
\author{Daniel Cartin}
\email{cartin@naps.edu}\affiliation{Naval Academy Preparatory School, 440 Meyerkord Avenue, Newport, Rhode Island 02841-1519}

\date{\today}
\begin{abstract}

At this point in time, there is very little empirical evidence on the likelihood of a space-faring species originating in the biosphere of a habitable world. However, there is a tension between the expectation that such a probability is relatively high (given our own origins on Earth), and the lack of any basis for believing the Solar System has ever been visited by an extraterrestrial colonization effort. This paper seeks to place upper limits on the probability of an interstellar civilization arising on a habitable planet in its stellar system, using a percolation model to simulate the progress of such a hypothetical civilization's colonization efforts in the local Solar neighborhood. To be as realistic as possible, the actual physical positions and characteristics of all stars within 40 parsecs of the Solar System are used as possible colony sites in the percolation process. If an interstellar civilization is very likely to have such colonization programs, and they can travel over large distances, then the upper bound on the likelihood of such a species arising per habitable world is on the order of $10^{-3}$; on the other hand, if civilizations are not prone to colonize their neighbors, or do not travel very far, then the upper limiting probability is much larger, even of order one.

\end{abstract}

\maketitle

\section{Introduction}
\label{intro}

After decades of human exploration of space, it appears that there have been no visitors to the Solar System from other star systems. This is in sharp contrast to the (perhaps naive) view that, with hundreds of billions of stars in the Milky Way galaxy, the likelihood of finding habitable planets around many of these stars, and the possibility of life arising on a good proportion of these planets -- as evidenced by the seeming rapidity of life developing on the young Earth -- there should be a large number of technological civilizations, and at least one of these should have developed spacecraft capable of reaching the Solar System. The variance between this supposition, and the lack of evidence for any visitations, is known as the Fermi paradox~\cite{Brin83, Hart75}. Its resolution has vexed many minds over the decades, although at this point, empirical data sufficient to address the question is lacking.

This has not stopped the development of theoretical models whose intent is to provide insight into the Fermi paradox. An as example appropriate to our discussion, Landis~\cite{Landis} originally developed a model of interstellar colonization processes as a simple percolation model, where each settled system has a given probability $p$ of colonizing all of its neighboring systems; since the model was simulated on a cubic lattice, this fixed the number of such neighbors to be six for every single star system. A generalization of this model was given by Hair and Hedman~\cite{HaiHed}, where the cubic lattice was still used, but colonization along diagonal connections was allowed  -- so each system had up to 26 neighboring systems -- and $p$ was the probability for each neighboring system (rather than all) to be colonized by a neighbor. However, neither model used the actual ``geography" of star systems; this aspect was included in the model developed by Cartin~\cite{Car14}. There a percolation model was also used, but instead of a fixed lattice, the observed positions of all known star systems within 40 parsecs (pc) of the Solar System were taken as colonization sites.

The main focus of Ref. 4 was the following question: {\it supposing each star system is the point of origin for a space-faring civilization with a given sociological drive and technological prowess, how many could reach the Solar System with their colonization process?} To codify this, the two parameters of the model were a probability $p$ for a given settled system to colonize a particular neighboring stellar system (``sociological drive''), and the maximum travel distance $D_{max}$ that its spacecraft can travel, which thus determines adjacency (``technological prowess''). The Monte Carlo simulations of these percolation processes allowed the calculation of the number of systems originating colonization programs that could reach the Solar System as a function $\phi(p, D_{max})$. In other words, $\phi(p, D_{max})$ is a correlation function reflecting whether the Solar System and any other given star system are in the same cluster of systems colonized by a given interstellar civilization.

This paper asks a different question, namely: {\it for this calculated function $\phi(p, D_{max})$ and the evident lack of visitors to the Solar System, what are the upper limits on the likelihood of spacefaring life originating on a given habitable planet, as a function of $p$ and $D_{max}$?} The number of interstellar civilizations reaching the Solar System must be the number of stellar origin sites for colonization efforts could possibly reach us, multiplied by the probability of a star-faring civilization arising at each star system. From the self-evident lack of visitors -- meaning the number of colonizing efforts reaching the Solar System is less than one -- the probability of a species willing and able to execute a colonization program must be simply the inverse of the probability of such civilizations evolving in a given star system. Section \ref{constraint} develops this idea further, by using an equation for the expected number of interstellar civilizations based on a break-down of the various relevant factors in a manner similar to the Drake equation, and solving for the probability of a spacefaring species arising on a given habitable world in terms of current constraints on such exoplanets. As with Ref. 4, all types of stellar systems are lumped into three broad categories -- bright single stars, dim single stars, and multiple star systems. Specific forms of $\phi_\alpha (p, D_{max})$ are obtained by fitting the results of the simulations in Ref. 4 to a modified power law, where $\alpha$ runs over the three system types. Using these fitting functions, an example case is shown for a choice of habitable planet occurrence rates around various star system types, and the maximum possible probability of a spacefaring species arising within the local Solar neighborhood are calculated as a function of $p$ and $D_{max}$. These results are summarized and discussed in Section \ref{conclusions}.

\section{Constraints on local interstellar civilizations}
\label{constraint}

As mentioned in Section \ref{intro}, this work begins with an equation based on the Drake equation, which parametrizes the number of extraterrestrial civilizations in the Milky Way galaxy from a combination of astrophysical, biological, technological and sociological inputs, with an eye towards estimating the total number of civilizations which are available for radio or other communications. A similar idea is useful in estimating the possibility of direct contact with an alien civilization. Specifically, such visits are only possible if (1) a potentially space-faring civilization arises in another star system, and (2) this civilization has the willingness and technology to reach the Solar System, possibly after colonizing one or more star systems along the way. It is now possible to make educated guesses on the likelihood of the first point. The result of this computation is a variable $N_k \in [0, 1]$, for a star system $k$, where $N_k$ denotes the probability of an interstellar civilization arising in star system $k$. Note that we use the term ``system'' here, since we include the possibility of space-faring species originating within a multiple star system. The second is an appropriate setting for the percolation model developed earlier in Ref. 4. The results calculated in that paper implicitly give a correlation function between the Solar System and all other systems in question, e.g. the probability that the Solar System and another given system are both within the same colonization cluster, based on a choice for colonization probability $p$ and travel distance $D_{max}$. For a stellar system $k$, this correlation probability is denoted $\phi_k (p, D_{max})$. Then the expected number $\langle N \rangle$ of civilizations visiting the Solar System is given by\footnote{Equations (2) and (3) of Brin~\cite{Brin83} are similar in spirit to the equations presented here, but use different parameters to define the model space.}
\begin{equation}
	\langle N \rangle = \sum_k N_k \phi_k (p, D_{max})
\end{equation}
We can parametrize $N_k$ much like the Drake equation to get
\begin{equation}
\label{Fermi-Drake-eqn}
	\langle N \rangle = \sum_k f_{p, k} n_{h, k} f_{\ell, k} f_{S, k} \phi_k (p, D_{max})
\end{equation}
where the parameters $f_p, n_h, f_\ell,$ and $f_S$, for each stellar system $k$, are defined in Table \ref{def}.

\begin{table}[hbt]
\begin{tabular}{cl}
Symbol 	& Definition						\\
\hline
$f_p$		& Fraction of stars with planets			\\
$n_h$		& Number of those planets within			\\
		& the habitable zone of their stellar system	\\
$f_\ell$	& Fraction of planets in habitable zone that	\\
		& develop life					\\
$f_S$	& Fraction of life-bearing planets developing		\\
		& a star-faring civilization				\\
\end{tabular}
\caption{\label{def}Definitions of parameters used in equation (\ref{Fermi-Drake-eqn}) for the expected number $\langle N \rangle$ of interstellar civilizations reaching the Solar System. The total probability $N_k$ of a space-faring civilization arising in a given star system is the product $N_k = f_p n_h f_\ell f_S$.}
\end{table} 

Since Ref. 4 only considers those star systems within 40 pc, we limit our discussion to that volume to find the expected number of civilizations reaching the Solar System. Theoretically, at least, the astrophysical properties of each stellar system within that volume -- i.e. the number of planets in each system, and the number of those within the habitable zone for that system -- could be directly measured, or at least estimated based on the age, spectral type, and metallicity of each star present in the system~\cite{WinFab}. However, even those projects such as the Kepler spacecraft have garnered a huge amount of information about exoplanets, the study of other worlds is still young, and only broad outlines can be seen at the moment. Thus, to match current knowledge, the sum (\ref{Fermi-Drake-eqn}) for the expected number of visiting civilizations is divided into three pieces, according to whether each system consists of a single star of spectral class K or earlier (``bright''), a single star of class M or other non-stellar object (``dim''), or a system of multiple stars (``multiple'').

In addition, there are two parameters more in the astrobiological realm than the astronomical, namely $f_\ell$ and $f_S$. The fraction $f_\ell$ of habitable zone worlds where life evolves is unknown at the moment\footnote{As noted by Spiegel and Turner~\cite{SpiTur}, current knowledge about the evolution of life (specifically on Earth) does little to limit the value of $f_\ell$.}, although this could be constrained by potential near-term space missions, such as those craft currently exploring Mars, or possible missions for Europa, Titan and Enceladus. On the other hand, the fraction $f_S$ of inhabited worlds where an interstellar civilization arises is completely unknown, since as far as we know, the answer is zero. In light of our ignorance of the processes behind $f_\ell$ and $f_S$, it is reasonable to keep the model as simple as possible; thus, for the following the product $f_\ell f_S$ is assumed to be the same for all star system types, and the remaining parameters are taken to be identical for a given system type $\alpha \in \{b, d, m\}$. This means the sum (\ref{Fermi-Drake-eqn}) for the expected number of visiting civilizations is given by
\[
	\langle N \rangle = (f_\ell f_S) \biggl[ \sum_\alpha (f_p n_h)_\alpha \sum_{k \in \alpha} \phi_k (p, D_{max}) \biggr]
\]
The results of Ref. 4 allows the functions
\[
	\langle N_\alpha (p, D_{max}) \rangle = \sum_{k \in \alpha} \phi_k (p, D_{max})
\]
to be calculated for all systems $k$ within 40 pc of the Solar System with a particular system type $\alpha$, so the previous equation (\ref{Fermi-Drake-eqn}) for $\langle N \rangle$ will be used in the form
\begin{equation}
\label{FD-eqn2}
	\langle N \rangle = (f_\ell f_S) \sum_{\alpha \in \{b, d, m\} } (f_p n_h)_\alpha \langle N_\alpha (p, D_{max}) \rangle
\end{equation}

This equation for the expected number of interstellar civilizations to reach the Solar System is not helpful as it stands, since it depends on unknown probabilities $f_\ell$ and $f_S$. However, the only observable fact that can be used in this discussion -- namely, the complete lack of evidence that the Solar System has been visited by extraterrestrial spacecraft -- does serve to constrain the product $f_\ell f_S$. In other words, the fact that $\langle N \rangle < 1$ allows the calculation of upper limits on the product $f_\ell f_S$, since this limit on the maximum value of the equation (\ref{Fermi-Drake-eqn}) for $\langle N \rangle$ implies
\begin{equation}
	(f_\ell f_S) < \biggl[ \sum_{\alpha \in \{b, d, m\} }  (f_p n_h)_\alpha \langle N_\alpha (p, D_{max}) \rangle \biggr]^{-1}
\end{equation}
All quantities on the right-hand side of this inequality are either probabilities now grounded in exoplanet surveys ($f_p$) and climate models ($n_h$) for these worlds, or else calculated from the percolation model ($\langle N_\alpha \rangle$) of Ref. 4, based on assumed values for the sociological drive and technological abilities of an interstellar civilization, i.e. the colonization probability $p$ and maximum travel distance $D_{max}$, respectively.

As a way of seeing the behavior of the number of systems that can reach the Solar System, we fit each of the three functions $\langle N_\alpha \rangle$ to a modified power law in $p$, given by
\begin{equation}
\label{fit-eqn}
	\langle N_\alpha (p, D_{max}) \rangle = A_\alpha p^{c_{1, \alpha} D^2 _{max} + c_{2, \alpha} D_{max} + c_{3, \alpha}}
\end{equation}
This gives us the coefficients given in Table \ref{coeff-table}, where the values of $\langle N_\alpha \rangle$ are fit for $p \in [0, 1]$ and $D_{max} \in [3.0, 7.0]$ pc. The large and negative coefficient of the linear $D_{max}$ terms in the exponent show that, as the maximum travel distance increases, a smaller value of the colonization probability is necessary in order to achieve the same overall value of $\langle N_\alpha \rangle$, for a given system type. However, the positive quadratic coefficient shows that this tendency levels off a bit as $D_{max}$ increases. Both of these effects are more pronounced for the bright stars and multiple star systems than they are for the dim stars.

\begin{table}[hbt]
\begin{tabular}{ccccc}
System type	& $A$		& $c_1$ (pc$^{-2}$)	& $c_2$ (pc$^{-1}$)	& $c_3$		\\
\hline
bright	& 2074	& 0.4905			& -5.743			& 18.52		\\
dim		& 1598	& 0.2351			& -2.714			& 9.668		\\
multiple	& 1575	& 0.4896			& -5.753			& 18.60		\\
\end{tabular}
\caption{\label{coeff-table}Fitting coefficients for the functions $\langle N_\alpha (p, D_{max}) \rangle$ giving the number of stellar systems of a given type $\alpha \in \{b, d, m\}$ able to reach the Solar System for a given choice of colonization probability $p$ and maximum travel distance $D_{max}$ (in pc). The functional form of $\langle N_\alpha (p, D_{max}) \rangle$ is given in equation (\ref{fit-eqn}).}
\end{table}

\begin{figure}
	\includegraphics[width = \columnwidth]{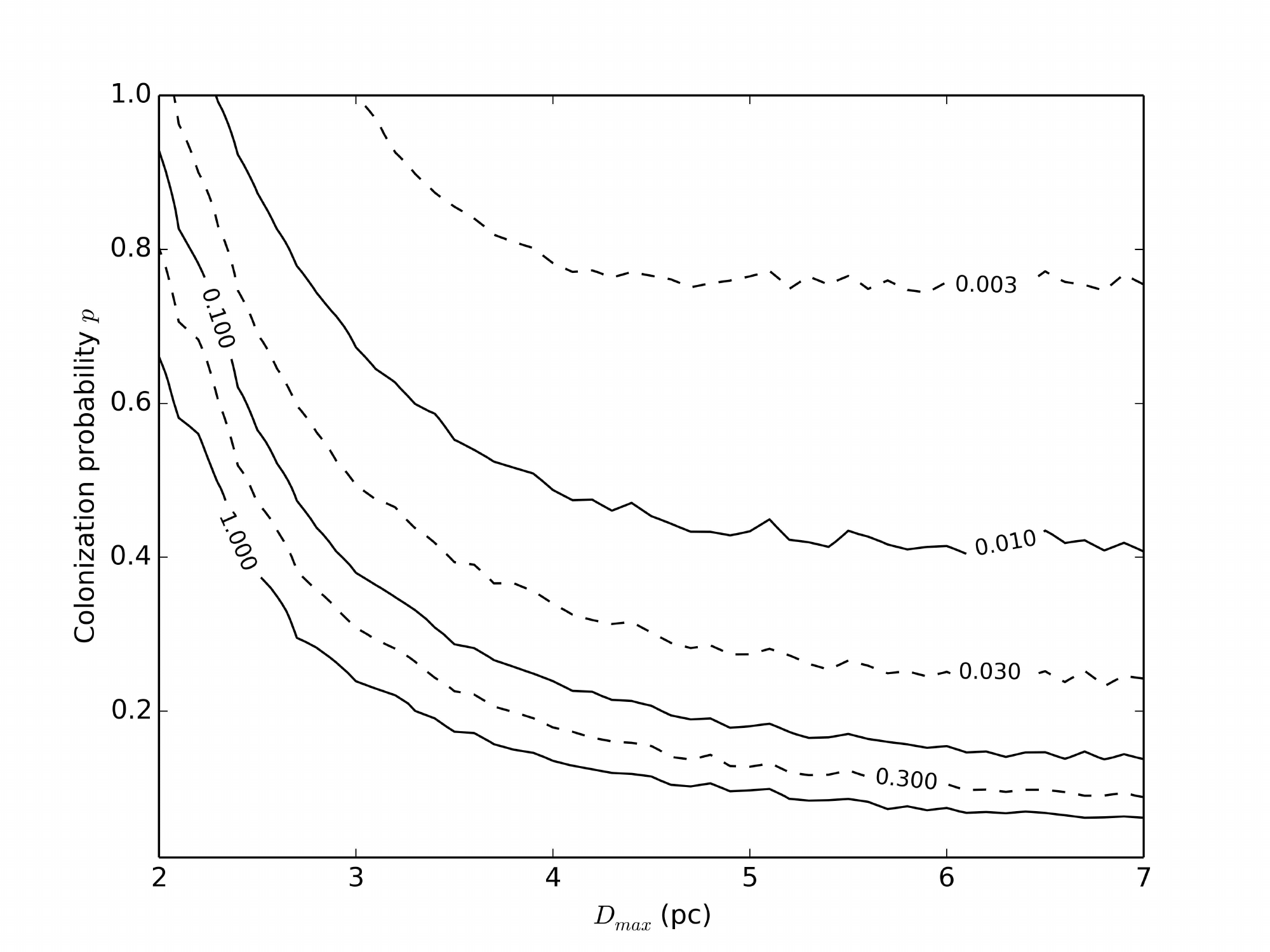}
	\caption{\label{bioProb}Contour plot of the maximum possible probability $f_\ell f_S$ of a star-faring civilization arising on an Earth-like planet in another star system, with spacecraft from that civilization reaching the Solar System. This graph uses the choice of Earth-like planets existing in other star systems $(f_p n_h)_\alpha = (0.0232, 0.309, 0.0174)$, for bright and dim stars, and multiple star systems, respectively.}
\end{figure}

As a specific instance of the use of this model, values for the parameters are chosen based on a geometric mean of results given in the literature, cited by Winn and Fabrycky~\cite{WinFab}. Specifically, the occurrence rates from Table 2 of that work are used, where the three listings for M stars are used for dim stars, the four entries for FGK are combined with the GK entry for bright stars, and the occurrence rate for habitable planets in multiple star systems is chosen as 75\% of that for bright stars. This gives the parameter choices of
\begin{eqnarray*}
	(f_p n_h)_\alpha &=& [ (f_p n_h)_{bright}, (f_p n_h)_{dim}, (f_p n_h)_{multiple}] 	\\
			&=& (0.0232, 0.309, 0.0174)
\end{eqnarray*}
The maximum possible values of $f_\ell f_S$ are then calculated, and a contour plot showing the results is given in Figure \ref{bioProb}. We now discuss the results shown in this figure; these comments relate specifically to the choice of $(f_p n_h)_\alpha$ used here, but the broad outline remains the same regardless of what values are chosen.

\begin{itemize}

	\item Interstellar civilizations capable of making longer journeys between star systems (higher $D_{max}$), and more likely to begin these journeys (larger $p$), have a higher chance of reaching the Solar System. Thus, the empirical fact that visiting spacecraft of this type are not observed places stronger limits on $f_\ell f_S$, and so the upper limit of this product decreases in size as we move towards the upper-right hand corner of Figure \ref{bioProb}.
	
	\item Specifically, the maximum range of travel distances for interstellar colonizers studied in this research was chosen to be 7.0 pc. If such a civilization always colonized its neighbors ($p = 1$), then the upper limit on the probability of such a species arising within a biosphere less than 40 pc away from the Solar System is $f_\ell f_S \le 0.00171$. In other words, at most only one of out every 585 habitable planets within the local Solar neighborhood could be the cradle of an interstellar civilization in order not to have evidence of their spacecraft in our Solar System.

	\item There is a relatively large regime of colonization probabilities $p$ and maximum travel distances $D_{max}$ where interstellar civilizations would never reach the Solar System, even if technological civilizations capable of star travel arose on every single Earth-like planet in the Solar neighborhood. The reach of these interstellar civilizations could be quite sizable, as seen in Ref. 4. However, the practical effect here is that there is a region in $p-D_{max}$ space where $\langle N \rangle < 1$ regardless of the value of $f_\ell f_S$.

\end{itemize}

Recall that the product $f_\ell f_S$ deals with the chances of an interstellar civilization arising on a given habitable planet, where we have estimated values for the expected number of such planets per stellar system (as divided between bright and dim stars, and multiple systems). Since the latter is on the order of 1-10\%, then the most stringent limit on the probability of a space-faring species evolving within any star system is around $10^{-5}$. Relative changes in the sizes of $(f_p n_h)_\alpha$ for the three different system types $\alpha$ may change the shapes of the contours given in Figure \ref{bioProb}, but the upper limit values for a given $p$ and $D_{max}$ will be altered only for absolute changes in the order of magnitude for the number of habitable planets per system, regardless of system type.

\section{Conclusions}
\label{conclusions}

In this paper, the model of interstellar colonization as a percolation process, originally developed by Landis and applied to the physical positions of all star systems within 40 pc, is used to develop upper limits on the probability of an interstellar civilization arising on an Earth-like planet, and spacecraft of that civilization reaching the Solar System. The model was parametrized in a manner similar to the Drake equation, with coefficients giving the fraction $f_p$ of stars with planets, and the number of those planets $n_h$ within the habitable zone of each stellar system, the fraction of habitable planets $f_\ell$ that actually develop life, and finally the fraction of those biospheres $f_S$ which originate a space-faring civilization. At this point in our knowledge, there are now empirical constraints on the first two parameters, but the latter are more or less free. The philosophy of this paper is to use a simple model in order to place some upper limits on those values dealing with the origin of life and interstellar civilizations, since there is currently no observational data on these issues. Our ignorance of many of the processes involved in the evolution of a technological species, and the likelihood of that species choosing to develop interstellar settlements leads to the choice of the simple model given here, with as few parameters as possible to give a reasonable look at the question at hand.

However, in some sense we have just moved our ignorance -- instead of leaving $f_\ell$ and $f_S$ open, this work puts upper limits on those values at the expense of two new coefficients, the probability $p$ of a space-faring species to colonize one of its neighboring star systems, and the maximum distance $D_{max}$ that can be considered "neighboring". Yet this shift in focus may be helpful, in that it puts a different light on questions relating to the Fermi paradox. Indeed, as shown elsewhere~\cite{Car10}, a maximum travel distance $D_{max}$ of about 3 pc is sufficient to allow effective interstellar travel, i.e. actual travel paths are less than twice the shortest distance between end points, with the benefit of passing through intermediate star systems, allowing for replenishment and refit. As for the colonization probability $p$, as yet humanity itself has not taken this step, but it is for each of us to evaluate the likelihood of this happening when technology has reached the point where this question is a viable one.

The percolation model presented here, however, only includes nodes representing all star systems with 40 pc of the Solar System, and thus does not account for the possibility of interstellar civilizations arising outside of that volume. Indeed, it is perfectly conceivable that a galaxy-wide colonization effort may have started from any point within the Milky Way, as long as the colonization probability $p$ is large enough, leading to much smaller upper bounds on $f_\ell f_S$. To truly study this issue, the percolation model would also have to be extended onto the galactic scale. The stellar number density -- the expected number of star systems found inside a cubic parsec of volume, for example -- is not constant over the Milky Way, but varies according to both distance away from the center of the galaxy, as well as scale height away from the galactic plane~\cite{SDSS}. Thus, the study of percolation clusters using a physically realistic distribution of simulated stellar systems would not correspond to previous models from statistical mechanics. Said in a different way, the critical probability $p_c$, i.e. the probability to have a galaxy-spanning cluster, needed as a function of $D_{max}$ is currently unknown. Thus, the size of the largest cluster, corresponding to the interstellar civilization with the greatest reach, as a function of the colonization probability $p$ and the maximum travel distance $D_{max}$ between two adjacent systems, remains a problem to be studied in future simulations. However, having a relatively low value of $p_c$ would imply that the upper limits on $f_\ell f_S$ would be very small, or else a galactic civilization would have arisen and easily reached out Solar System, contrary to observation.

Another important caveat for the work here is that none of the relative motion of star systems in the Solar neighborhood is taken into account. This motion leads to changes in positioning over periods of only $10^3 - 10^4$ yr~\cite{BailerJones}. The Monte Carlo simulations used to model the percolation process in Ref. 4 are ``timeless'', in the sense that all systems are static, and the speed of colonizing spacecraft is not considered. It is likely this time scale is roughly the same as the process for a newly settled star system from an interstellar civilization to built its own colonizing spacecraft, and then this craft reaching a new system, so the dynamical changing of available star systems for a star-faring species would definitely affect the results presented here; this also is an issue for further work.

\end{document}